\documentclass[twocolumn,showpacs,preprintnumbers,amsmath,amssymb]{revtex4}

\usepackage{dcolumn}
\usepackage{bm}
\usepackage{epsfig}

\begin{document}

\title{Electrokinetic-flow-induced  viscous drag on a tethered DNA inside a nanopore}
\author{Sandip Ghosal}
\affiliation{%
Northwestern University, Department of Mechanical Engineering\\
2145 Sheridan Road, Evanston, IL 60208
}%
\date{\today}
\begin{abstract}
Recent work has shown that the resistive force arising from  viscous effects within the pore region could explain [Ghosal, S.  Phys. Rev. E. {\bf 71}, 051904 (2006) \& Phys. Rev. Lett. {\bf 98}, 238104 (2007)]
observed translocation times in certain experiments involving voltage driven translocations of DNA through nanopores. 
The electrokinetic flow inside the pore and the accompanying viscous effects also play a crucial role 
in the interpretation of experiments where the DNA is immobilized inside a nanopore
[U. Keyser, {\it et al.} {\it Nature Physics} {\bf 2}, 473 (2006)].
In this paper the viscous force is explicitly calculated for a nanopore of cylindrical geometry.  It is 
found that the reduction of the tether force due to viscous drag and due to charge reduction by Manning 
condensation are of similar size.
The result is of importance in the interpretation of experimental data 
on tethered DNA.
\end{abstract}

\maketitle
The interaction of charged polymers, such as DNA with nanometer sized natural and artificial pores have received 
considerable attention recently~\cite{kasianowicz_PNAS96,meller_etal_PNAS00,ssdna_sequence_nbt,storm_nature,storm_physRevE05,lubensky_nelson,ghosal_PRE06}. 
Such studies are partly motivated by the desire to understand how polymers cross 
internal membranes of cells
~\cite{Alberts}. The possibility of developing devices capable 
of detecting properties of biopolymers at the single molecule level for applications such as rapid DNA  sequencing ~\cite{deamer_trendsinbiotech} 
is also a motivating factor for such studies.

\begin{figure}[t]
\center{
    \includegraphics[angle=0,width=3.5truein]{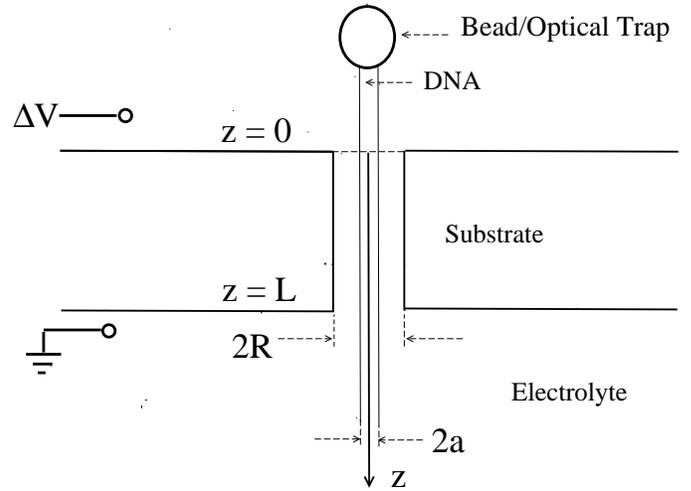}
    \caption{Sketch illustrating the tethered nanopore experiment with a cylindrical pore}
    \label{fig:geom}
    }
\end{figure}

In a recent paper Keyser {\em et al.} reported \cite{keyser_nature_phys06} experimental measurements in which a single strand of dsDNA 
was immobilized while threaded inside a nanopore by the application of a pulling force to counteract 
the electrical force on the DNA. This was achieved by attaching one end of the DNA to a Streptavidin coated 
polystyrene bead and holding the bead in a laser optical trap. The displacement of the bead from its equilibrium position 
 could be detected and used to measure the pulling force on the DNA. 
The measured value was found to be about 75 \% of the maximum electric force on the DNA within the pore 
 based on its bare charge irrespective of the electrolyte (KCl) concentration. This pulling force is however 
 determined by a complex interplay between electric forces and hydrodynamics, as noted by 
 Keyser {\em et al.}~\cite{keyser_nature_phys06}.
 The point of this calculation on an idealized physical model is to understand the relative importance of hydrodynamics 
 and the reduction of effective charge on the DNA due to Manning condensation in determining the observed pulling force.
  Since the DNA  as well as the internal walls of the pore are charged, the pore region
 has a cylindrically symmetric distribution of  oppositely charged counter-ions. In the 
presence of a strong electric field an electroosmotic flow~\cite{probstein} is therefore generated in this region that flows in a direction 
opposite to the direction in which the DNA would move if it were not immobilized (Figure~\ref{fig:geom}). This flow produces a 
hydrodynamic drag  on the DNA partially balancing the applied electrical force. In this paper, a simplified geometry of the pore 
region is used to calculate explicitly the viscous drag. 
It is shown that the  drag  is a significant fraction of the total force acting on the DNA and needs to be taken into 
account for a proper interpretation of experimental data on DNA nanopore interactions.

A simplified model is adopted in which the nanopore is regarded as a cylinder of radius $R$ (5.0 nm) 
and length $L$ (60 nm). The part of the DNA inside the nanopore is regarded as a uniformly charged 
cylinder of radius $a$ (1.1 nm) along the axis of the pore. The DNA has a linear charge density $\lambda$ (2 electron
charges every 0.34 nm -- the distance between adjacent bases) and a lower ``effective'' charge density of $\lambda_{e} = \lambda/q_{B}$
due to the Oosawa-Manning~\cite{Oosawa, Manning}  phenomenon of counter-ion condensation on its surface.
The factor $q_{B}$ is the Oosawa-Manning factor, it has the value of $q_{B}= 4.2$ for an ideal model 
of an infinite line charge in an unbounded electrolyte. Referring to the system sketched in 
Figure~\ref{fig:geom}, the fluid velocity in the pore is axially directed and is described by some function $u(r)$ where $r$ is the 
distance from the central axis. The electric potential 
is $-E_{0} z + \phi(r)$ where the first term is due to the externally applied axial electric field $E_{0}$ along 
the pore (the $z$-direction). The functions $u$ and $\phi$ are governed by the Stokes equation for viscous flow (with zero pressure gradient and an 
electric body force term) and the 
Poisson equation of electrostatics respectively:
\begin{eqnarray} 
\mu \frac{1}{r} \frac{d}{dr} \left( r \frac{d u}{d r} \right)   + \rho_{e} (r) E_{0} &=& 0 \label{eq:stokes}\\
\epsilon   \frac{1}{r} \frac{d}{dr} \left( r \frac{d \phi }{d r} \right)&=& - \rho_{e} \label{eq:poisson}
\end{eqnarray} 
where $\epsilon$ is the permittivity of the electrolyte and $\rho_{e}$ is the electric charge density due to ions.
The classical  boundary conditions of `no slip' are assumed for the velocity: 
\begin{equation} 
u(a) = u(R) = 0.
\label{eq:noslip}
\end{equation} 
Eliminating $\rho_{e}$ from the pair of equations (\ref{eq:stokes}) and (\ref{eq:poisson})  and using (\ref{eq:noslip}) determines 
$u(r)$ in terms of the potential $\phi(r)$:
\begin{equation} 
u(r) = \frac{\epsilon E_{0}}{\mu}  \left[ \phi(r)-\phi(R)  + \Delta \phi \,  \frac{\ln (r/R) }{\ln (a/R) } \right]
\end{equation} 
where $\Delta \phi = \phi(R)-\phi(a)$.
The viscous force (along the $z$-axis) on the DNA is then
\begin{equation} 
F_{v} = 2 \pi a L \mu u^{\prime}(a) =  2 \pi a \epsilon E_{0} L \left[ \phi^{\prime}(a) + \frac{\Delta \phi}{ a \ln (a/R) } \right].
  \label{eq:F_visc}
\end{equation} 
If $S$ is the surface charge density on the channel wall, then by Gauss' law,
\begin{eqnarray} 
- 2 \pi a \epsilon  \phi^{\prime}(a) &=& \lambda_{e} \label{eq:bc_in}\\
 \epsilon  \phi^{\prime}(R) &=& S 
 \label{eq:bc_out}
\end{eqnarray} 
Using the first of these equations to eliminate $\phi^{\prime}(a)$ and noting that the electrical 
force on the DNA, $F_{e} = \lambda_{e} L E_{0}$, 
equation~(\ref{eq:F_visc}) may be written as 
\begin{equation} 
- \frac{F_{t}}{F_{e}} = \frac{F_{e}+F_{v}}{F_{e}} =  \frac{2 \pi \epsilon \Delta \phi}{\lambda_{e} \ln (a/R) } 
\label{eq:tetherforce}
\end{equation} 
where $F_{t}=-F_{v}-F_{e}$ is the tether force. 
\begin{figure}[t]
\begin{center}
        \includegraphics[angle=0,width=3.5truein]{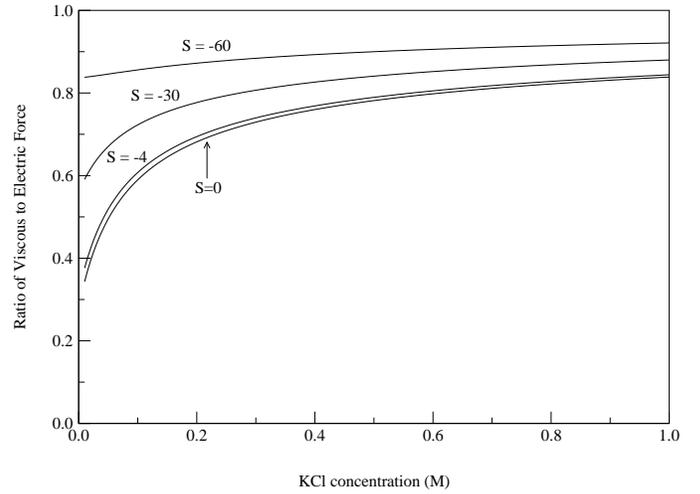}
    \caption{Ratio of viscous to electric force on DNA ($-F_{v}/F_{e}$) plotted as a function 
    of KCl concentration assuming a charge of 2 electrons per base pair on the DNA (Manning factor 
    of unity) and a constant surface charge on the pore wall parametrized by the surface charge density $S$
    (in mC/${\rm m}^{2}$).}
    \label{fig:force}
\end{center}
\end{figure}

In order to calculate the ratio $F_{v}/F_{e}$ from equation~(\ref{eq:tetherforce}), all that remains to be done 
is to calculate the quantity $\Delta \phi$. 
If the potential $\phi$ anywhere in the gap does not greatly exceed
$k_{B}T/e$ (about 30 mV at room temperature; $k_{B}$ is the Boltzmann factor, $T$ the absolute temperature and $e$ the magnitude of electric charge) then $\phi$ may be computed from the Debye-H\"{u}ckel model ($\lambda_{D}$ is the Debye length):
\begin{equation} 
 \frac{1}{r} \frac{d}{dr} \left( r \frac{d \phi}{d r} \right)   = \frac{\phi}{ \lambda_{D}^{2} },
 \label{eq:debye_huckel}
\end{equation} 
with the boundary conditions (\ref{eq:bc_in}) and (\ref{eq:bc_out}). 
The solution to that problem is 
\begin{equation} 
\phi(r) = \frac{\lambda_{e}}{2 \pi \epsilon}  \left[ A I_{0} \left( \frac{r}{\lambda_{D}} \right)  + B K_{0} \left( \frac{r}{\lambda_{D}} \right) \right] 
\label{eq:soln_bessel}
\end{equation} 
where the constants $A$ and $B$ 
may be compactly expressed in terms of the following dimensionless variables 
$a_{*} = a / \lambda_{D}$, $R_{*}=R/\lambda_{D}$ 
and $S_{*}= ( 2 \pi  a S )/ \lambda_{e}$. Thus, 
\begin{eqnarray} 
A &=& \frac{ S_{*} K_{1}(a_{*}) + K_{1} (R_{*}) }{a_{*} [ I_{1}(R_{*}) K_{1}(a_{*}) - I_{1} (a_{*}) K_{1} (R_{*}) ] } \label{eq:coeff_A}\\
B &=& \frac{ S_{*} I_{1} (a_{*}) +  I_{1}(R_{*}) }{a_{*} [ I_{1}(R_{*}) K_{1}(a_{*}) - I_{1} (a_{*}) K_{1} (R_{*}) ] }  \label{eq:coeff_B}
\end{eqnarray}
where $K_{n}$, $I_{n}$ ($n$ is a non-negative integer) are modified Bessel functions of integral order. 
The potential drop $\Delta \phi$ in equation (\ref{eq:tetherforce}) is now easily found from 
equation~(\ref{eq:soln_bessel}) 
\begin{equation} 
\Delta \phi = \frac{\lambda_{e}}{2 \pi \epsilon}  \left[ A \Delta I_{0} + B \Delta K_{0} \right] 
\end{equation} 
where $\Delta I_{0} = I_{0} ( R_{*} ) - I_{0}(a_{*})$ and $\Delta K_{0} = K_{0} (R_{*}) - K_{0}(a_{*})$.

The result of the calculation as described above is shown in Figures~\ref{fig:force}.
The figure shows the magnitude of the viscous to the electric force ($-F_{v}/F_{e}$) 
evaluated from equation~(\ref{eq:tetherforce}) using the value of $\lambda_{D}$ appropriate 
 for a symmetric binary electrolyte~\cite{probstein}. 
The effective charge of the DNA is assumed to be the same as the bare charge ($\lambda_{e} = \lambda$) 
of -2e per base pair and the applied Voltage 
is $\Delta V = -120$ mV. The surface charge concentration on the substrate, $S$ has been assumed 
independent of the KCl concentration. From measurements of conductance and streaming 
potentials it has been shown~\cite{stein_kruithof_dekker_prl04,dekker_nano_lett06} that 
in Si/Si${\rm O}_{2}$ nanopores $S \approx - 60$   mC/${\rm m}^{2}$
for KCl concentrations greater than about $0.1$ M. However, for low concentrations, 
the surface charge density drops substantially and needs to be calculated from a more 
elaborate model that takes into account the equilibrium of surface reactions  at the interface. For very 
low concentrations, $S \approx - 4$ mC/${\rm m}^{2}$. Due to the approximate nature of 
our model  it does not seem worthwhile 
to attempt to incorporate the proper dependence of $S$ on KCl concentration. Instead, it suffices 
to show how $-F_{v}/F_{e}$ varies with salt concentration for several fixed values of 
$S$ between $- 4$ to $- 60$ mC/${\rm m}^{2}$ as shown in Figure~\ref{fig:force}. The important 
feature that these curves illustrate is that $-F_{v}/F_{e}$ is essentially constant for 
most of the KCl concentration range at a value of around 0.7 -- 0.8. Thus, the viscous 
force is not small, and furthermore, if the viscous force were neglected  and the reduction 
in the electrical pulling force were  attributed to counter-ion condensation, it would appear that the
DNA effective charge is lowered by about $75$ percent (just the right amount to lead one 
to conclude that charge reduction by the Manning factor of  $q_{B}=4.2$  is being validated!).
Figure~\ref{fig:force} shows an increase in the hydrodynamic drag with increase in the magnitude of $S$, 
 because the surface charge on the pore walls enhance the electroosmotic flow due to the DNA charge.
 A model that properly accounts for the variation of $S$ with KCl concentration
is expected to follow the $S = - 4$ mC/${\rm m}^{2}$ curve closely for low concentrations 
(below about $0.1$ M) and asymptote to the  $S = - 60$ mC/${\rm m}^{2}$ curve at high concentrations.
The distribution of counter ions in the  calculations presented here was treated by means of the equilibrium 
Debye-H\"{u}ckel theory and one may question whether that corresponds to the experimental conditions. 
Taking the ratio $-F_{v}/F_{e} \approx 0.75$, equation~(\ref{eq:tetherforce}) gives $\Delta \phi \approx 80$ mV
if for $\lambda_{e}$ one assumes the DNA bare charge of two electronic charges per base pair. If 
this is reduced by the Manning factor of $q_{B}=4.2$, then $\Delta \phi \approx 19$ mV. Though the 
formal requirement for the Debye-H\"{u}ckel linearization is $|\phi| << k_{B}T/e \approx 33$ mV, in practice 
the double layer structure calculated from the Debye-H\"{u}ckel theory does not deviate substantially from 
the more accurate Poisson-Boltzmann calculation as long as the maximum value of  $|\phi|$ is not 
substantially larger than $2 k_{B} T/ e \approx 66$ mV~\cite{hunter}. Therefore the Debye-H\"{u}ckel theory certainly 
suffices for our present purpose. For the cylindrical geometry considered here, the applied potential 
does not disturb the equilibrium Debye layer structure, since the applied field is always along the iso-concentration 
 surfaces of the ions. However, for the real nanopore, the applied electric field may have a radial 
 component, and one may ask if this is strong enough to distort the equilibrium Debye layer. Since 
 $|\Delta V| = 120$ mV and $L = 60$ nm, this imposed field is $E_{ext} = |\Delta V|/L \approx 2 \times 10^{6}$ V/m. 
 The radial field within the Debye layer may be estimated as $E_{int} = | \Delta \phi | /(R - a) \approx 5  \times 10^{6}$ V/m
 with Manning condensation and $E_{int} \approx 20 \times 10^{6}$ V/m with the DNA bare charge. Thus, 
 though the distortion of the double layer can be neglected for the purpose of obtaining a rough estimate, 
 it should be accounted for if one desires an accurate calculation of the viscous force. In order to do so, 
 the cross-sectional shape of the nanopore must be known.

Analysis of this simplified model suggests that 
a more careful modeling is needed in order to properly interpret the 
 Keyser {\it et al.}~\cite{keyser_nature_phys06} experiments.
Such a model  should account for hydrodynamic drag while 
taking into account the proper pore shape, the variation of substrate charge with KCl, possible departures of the 
equilibrium potential from the Debye-H\"{u}ckel model and other relevant conditions 
of the experimental set up. Numerical simulation on a more elaborate model incorporating these details 
used in conjunction with the experimental data could provide a more complete picture of the effective charge on DNA 
inside a nanopore. One may be tempted to question the use of classical continuum hydrodynamics to 
flows on the nanometer scale. However, the classical approach has already been shown to give 
results in reasonable agreement with experiments on DNA translocation through solid state nanopores~\cite{ghosal_PRE06,ghosal_PRL07}.
Molecular dynamic simulations, such as those presented by Aksimentiev {\it et al.}~\cite{aksi_bpj04} 
could be used to further refine these calculations and to show that the effects described here persist even if continuum hydrodynamics 
is replaced by a discrete molecular model. 
In situations where the length of the DNA polymer is much greater than the 
length of the nanopore, entropic forces due to random coiling of the polymer become significant. 
Such entropic effects  have been considered by Muthukumar~\cite{muthukumar99,muthukumar01,muthukumar03}.
In the limit $R \gg a$, the hydrodynamic friction with the pore walls becomes unimportant and the problem 
becomes one of determining the electric field that would immobilize a polyelectrolyte acted upon by given non-electrical forces 
in the presence Brownian fluctuations. This problem has been studied in its general form by 
Long {\it et al.}~\cite{long_viovy_ajdari_prl96,long_viovy_ajdari_biopolymers96}.
 
\bibliographystyle{prsty}

\begin{thebibliography}{10}

\bibitem{kasianowicz_PNAS96}
J. Kasianowicz, E. Brandin, D. Branton, and D. Deamer, Proc. Natl. Acad. Sci.
  {\bf 93},  13770  (1996).

\bibitem{meller_etal_PNAS00}
A. Meller {\it et~al.}, Proc. Natl. Acad. Sci. {\bf 97},  1079  (2000).

\bibitem{ssdna_sequence_nbt}
W. Vercoutere {\it et~al.}, Nature Biotechnology {\bf 19},  248  (2001).

\bibitem{storm_nature}
A. Storm {\it et~al.}, Nature Materials {\bf 2},  537  (2003).

\bibitem{storm_physRevE05}
A. Storm, J. Chen, H. Zandbergen, and C. Dekker, Phys. Rev. E {\bf 71},  051903
   (2005).

\bibitem{lubensky_nelson}
D. Lubensky and D. Nelson, Biophys. J. {\bf 77},  1824  (1999).

\bibitem{ghosal_PRE06}
S. Ghosal, Phys. Rev. E {\bf 74},  041901  (2006).

\bibitem{Alberts}
B. Alberts {\it et~al.}, {\em Molecular Biology of the Cell} (Garland
  Publishing, Taylor \& Francis Group, New York, U.S.A., 1994).

\bibitem{deamer_trendsinbiotech}
D. Deamer and M. Akeson, Trends in Biotech. {\bf 18},  147  (2000).

\bibitem{keyser_nature_phys06}
U. Keyser {\it et~al.}, Nature Physics {\bf 2},  473  (2006).

\bibitem{probstein}
R. Probstein, {\em Physicochemical Hydrodynamics} (John Wiley and Sons, Inc.,
  New York, U.S.A., 1994).

\bibitem{Oosawa}
F. Oosawa, {\em Polyelectrolytes} (Marcel Dekker, New York, U.S.A., 1971).

\bibitem{Manning}
G. Manning, J. Chem. Phys. {\bf 51},  924  (1969).

\bibitem{stein_kruithof_dekker_prl04}
D. Stein, M. Kruithof, and C. Dekker, Phys. Rev. Lett. {\bf 93},  035901
  (2004).

\bibitem{dekker_nano_lett06}
M. Smeets {\it et~al.}, Nano Letters {\bf 6},  89  (2006).

\bibitem{hunter}
R. Hunter, {\em Introduction to modern Colloid science} (Oxford University
  Press, Oxford. U.K., 2003).

\bibitem{ghosal_PRL07}
S. Ghosal, Phys. Rev. Lett. {\bf 98},  238104  (2007).

\bibitem{aksi_bpj04}
A. Aksimentiev, J. Heng, G. Timp, and K. Schulten, Biophys. J. {\bf 87},
  2086Ð2097  (2004).

\bibitem{muthukumar99}
M. Muthukumar, J. Chem. Phys. {\bf 111},  10371  (1999).

\bibitem{muthukumar01}
M. Muthukumar, Phys. Rev. Lett. {\bf 86},  3188  (2001).

\bibitem{muthukumar03}
M. Muthukumar, J. Chem. Phys. {\bf 118},  5174  (2003).

\bibitem{long_viovy_ajdari_prl96}
D. Long, J. Viovy, and A. Ajdari, Phys. Rev. Lett. {\bf 76},  3858  (1996).

\bibitem{long_viovy_ajdari_biopolymers96}
D. Long, J. Viovy, and A. Ajdari, Biopoloymers {\bf 39},  755  (1996).

\end{thebibliography}

\end{document}